\documentclass[a4paper,12pt]{article}

\begin{document}
\makeatletter
\renewcommand{\theequation}{\thesection.\arabic{equation}}
\@addtoreset{equation}{section}
\makeatother

\title{A Short Essay on Quantum Black Holes and Underlying Noncommutative  Quantized Space-Time }

\author{\large Sho Tanaka\footnote{Em. Professor of Kyoto University, E-mail: st-desc@kyoto.zaq.ne.jp }
\\[8 pt]
 Kurodani 33-4, Sakyo-ku, Kyoto 606-8331, Japan}
\date{}
\maketitle

\vspace{-10cm}
\rightline{}
\vspace{10cm}

\abstract{We emphasize the importance of noncommutative geometry or Lorenz-covariant quantized space-time towards ultimate theory of quantum gravity and Planck scale physics. We focus our attention on the statistical and substantial understanding of Bekenstein-Hawking's Area-Entropy Law of black holes  in terms of Kinematical Holographic Relation (KHR). KHR manifestly holds in Yang's quantized space-time as the result of kinematical reduction of spatial degrees of freedom caused by its own nature of noncommutative geometry and plays  an important role in our  approach without any recourse to the familiar hypothesis, so-called Holographic Principle. In the present paper, we  find out a {\it unified} form of KHR applicable to the whole region ranging from macroscopic to microscopic scales in spatial dimension $ d=3.$  We notice  a possibility of nontrivial modification of  Area-Entropy Law of Black Holes which becomes most remarkable in the extremely microscopic system close to Planck scale.

\newpage
\section{\normalsize Introduction} 

In our preceding paper, ``Where does Black-Hole Entropy Lie? - Some Remarks on Area-Entropy Law, Holographic Principle and Noncommutative Space-Time" (2014) [1], hereafter referred as I, we emphasized the importance of underlying noncommutative geometry or quantized space-time such as Snyder's  and Yang's Lorentz-covariant quantized space-time [2-5] towards ultimate theory of quantum gravity and Planck scale physics. We focused there most importantly our attention on the {\it statistical} and {\it substantial} understanding of Bekenstein-Hawking's Area-Entropy Law of black holes[6-9]  in terms of Kinematical Holographic Relation ({\bf KHR}) [10]. Indeed, as will be simply reviewed in scct.2, {\bf KHR} manifestly holds in Yang's quantized space-time as the result of kinematical reduction of spatial degrees of freedom caused by its own nature of noncommutative geometry and plays  an important role in our  approach, without any recourse to the familiar hypothesis, so-called Holographic Principle.

 In the present paper, first of all, we find out in sect.3 the following important fact that {\bf KHR} given in I gets a simple  and  {\it unified} form equally applicable to the whole regions ranging from macroscopic to microscopic. As the result, the {\it new} form of {\bf KHR} enables us to reconsider Area-entropy law of black holes all over the regions from macroscopic to extremely microscopic (see sect 4) . And finally in sect.5  we notice an  important possibility of the {\it nontrivial modification} of Area-Entropy Law of Black Holes, which becomes remarkable in the extremely microscopic black holes close to Planck scale.

The present paper is organized as follows. In sect.2, we first review the derivation of the {\it approximate} form of {\bf KHR} mentioned above for the subsequent arguments. Sect.3 is devoted to our central concern in our present research, that is, in subsect.3.1 we notice first of all  the {\it unified} form {\bf KHR}: ${\rm n_ {dof}} (V_3^L) =([L/\lambda]+1)^2$ in place of the approximate form of {\bf KHR} and in subsect.3.2 we statistically derive the entropy and mass of  $D_0$ brane gas systems  ranging from macroscopic- to microscopic- scales in d=3, under the full use of the novel {\bf KHR} mentioned above. Sect.4 is devoted to Schwarzschild black holes ranging from macroscopic to extremely microscopic scales in d=3, noticing there the existence and different behavior of two kinds of temperatures, $ {\it T_{H.R.}}$ and $T_S$ of black holes in both regions.  In the final sect.5, "Concluding Arguments and Further Outlook," first we give the supplementary and summarizing arguments on two kinds of temperatures of black holes (subsect.5.1) and notice the possible limits of applicability of area-entropy law of black holes (subsect.5.2).  Appendix A is devoted to the review of Yang's Quantized space-time and Appendix B to the historical background of  noncommutative quantized space-time, recollecting the related works  by W.K.~Heisenberg,  P.A.M.~Dirac and H.~Yukawa. 
  
\section{\normalsize Review of Kinematical Holographic Relation ({\bf KHR}) and Area-Entropy Law of Black Holes}
Let us briefly review here the derivation of  {\bf  KHR} for the subsequent arguments.  In I,  we started with the Kinematical Holographic Relation ( {\bf KHR} ) mainly for the {\it Macroscopic} system} given in the following form 
\begin{eqnarray}
 {\rm {\bf  KHR}}  \hspace{1cm}   n_{\rm dof} (V_d^L)= {\cal A} (V_d^L) / f_d,
\end{eqnarray}
that is, the proportional relation between $n_{\rm dof}(V_d^L)$ and ${\cal A} (V_d^L)$ with proportional constant $f_d, $\footnote { We use hereafter $f_d$ instead of  the misleading notation $G_d$ used in I.} where $n_{\rm dof} (V_d^L)$ and ${\cal A}(V_d^L)$, respectively, denote the number of degrees of freedom of any $d$ dimensional bounded spatial region  $V_d^L$ with radius $L$ in Yang's quantized space-time, and the boundary area of  $V_d^L$ in unit of Planck length.

 First, the region  $V_d^L$ is defined on any d-dimensional quantized spatial coordinate operators,  
 \begin{eqnarray}  
{{\hat X}_1}^2 + {{\hat X}_2}^2 + \cdots + {{\hat X}_d}^2 = L^{2}.
\end{eqnarray}

As was shown in detail in I (section 3), the most important concept,  the number of degrees of freedom of $V_d^L$,  $n_{dof} (V_d^L)$ in Eq.(2.1),  is found in a certain irreducible representation of $SO(d+1)$, a minimum subalgebra of Yang quantized space-time, which includes the above  $d$ spatial coordinate operators, $\hat{X}_1, \hat{X}_2, \cdots, \hat{X}_d$ in Eq.(2.2) needed to properly describe $V_d^L$, and is really constructed by the generators $\hat{\Sigma}_{M N}$ with $M, N$ ranging over $a,1,2, \cdots d.$( See Appendix A) Let us denote the irreducible representation by $\rho_l (V_d^L)$ with the characteristic integer $l$ which indicates the maximal eigenvalue of any generators,  $\hat{\Sigma}_{MN}$ of $SO(d+1).$ Then, ${\rm n_{dof}} (V_d^L)$ is reasonably identified with the {\it dimension} of $\rho_{[L/\lambda]}$, that is, 
$ {\rm n_{ dof}} (V_d^L) = {\rm dim} ( \rho_{[L/\lambda]} (V_d^L)).$

 According to the Weyl dimension formula applied to the irreducible representation of  $SO(d+1)$, the dimension of $\rho_l$ is given by

\begin{eqnarray}
{\rm dim}(\rho_l) = {l +\nu \over \nu}\  {(l+2\nu -1) ! \over l !( 2\nu - 1)!},
\end{eqnarray}
with $ \nu = (d-1)/2$ in the case $ d\geq 2.$ (see, more in detail [10])   

One immediately finds that 
\begin{eqnarray}
{\rm n_ {dof}}\ (V_d^L) &&= {\rm dim}\ ( \rho_{[L/\lambda]}\ (V_d^L))
 \nonumber \\
&& = {2[L/\lambda]+d-1 \over [L/\lambda]} {([L/\lambda] +d-2)! \over ([L/\lambda]-1)! (d-1)!} \nonumber \\     
&&\sim{ 2 \over (d-1) !} [L/\lambda]^{d-1} ,
\end{eqnarray}
where $[L/\lambda]$ denotes the nearest integer of $L/\lambda$ and $\lambda$ the short scale parameter in Yang's quantized space time (see Appendix A) and  identified with Planck length $l_P  ( = [G \hbar / c^3]^{1/2} )$ in I and in what follows. The expression in the last line holds for macroscopic system with $ [L/\lambda] \gg d,$ which was considered in I.

On the other hand, the boundary area of $V_d^L$ in the unit of $\lambda$,  ${\cal A} (V_d^L)$ is given by $ S^{d-1}$ with radius $ L/\lambda $, that is.
 
\begin{eqnarray}
\hspace{-1cm}{\cal A} (V_d^L) ={(2 \pi)^{d/2} \over {(d-2)!!}} (L/\lambda)^{d-1} 
&& {\rm for\ {\it d}\ even}
\nonumber\\                   =2 {(2\pi)^{(d-1)/2} \over {(d-2)!!}} (L/\lambda)^{d-1} &&{\rm for\ {\it d}\ odd}.
\end{eqnarray}

Comparing both Eqs. (2.4) and (2.5), one finally arrive at {\bf KHR} (2.1) with $f_d$ given by

\begin{eqnarray}
f_d  &&\sim\   {(2 \pi)^{d/2} \over 2} (d-1)!! \qquad  for\  d\  even \nonumber\\
      &&\sim\  (2 \pi)^{(d-1)/2} (d-1)! !\quad     for\   d\  odd
\end{eqnarray}
for the  {\it macroscopic} system with  $ [L/\lambda] \gg d,$ without any recourse to the familiar hypothesis, so-called Holographic Principle  (see, for instance, [20-22] ).

As was emphasized in I, the spatial structure of $V_d^L$ is described through some specific representation $ \rho_{[L/\lambda]} (V_d^L)$. Let us denote its orthogonal basis-vector system in the representation space, which we called Hilbert space I, as follows 

\begin{eqnarray} \rho_{[L/\lambda]}\ (V_d^L): \quad |\ m\ \rangle,  \qquad  m= 1,2,\cdots, n_{\rm dof}(V_d^L).    
\end{eqnarray}
The labeling number $m$ of basis vectors in ``Hilbert space I," plays the role of classical {\it spatial coordinates} of the classical space inside  $V_3^L$ and we called the  point  $[site]$ or $[site\ m].$ 

It is easy to imagine that {\bf KHR} Eq.(2.1) strongly suggests that the entropy of any statistical system realized in the spatial region  $V_d^L$  must be proportional not to the classical volume of $V_d^L$, but to the degrees of freedoms $n_{\rm dof} (V_d^L)  (= {\cal A} (V_d^L) / f_d)$, namely, it yields  a new  Area-Entropy Law. Indeed, in I, we derived the following form of a new {\it area-entropy relation}  of black hole in a purely {\it statistical} way, through a simple $D_0$ brane gas model constructed in Yang's quantized space time 
\begin{eqnarray}
S_S(V_3^{R_S}) = n_{\rm dof}(V_3^{R_S}) S_S[site].   
\end{eqnarray}

Indeed, the relation was shown by Eq. (53) in I, and in the related argument we concluded  that  $``S_S [site]$ represents a kind of {\it universal} unit of entropy of black holes, which appears as the entropy realized on each individual [{site] in any black hole, by taking a proper specific value,
\begin{eqnarray}
 S_S[site] = 4 \pi \eta 
\end{eqnarray}
and thus
\begin{eqnarray}
S_S [site] =  \pi  
\end{eqnarray}
under Bekenstein parameter $\eta = 1/4."$

In the present paper, the above argument given in I focusing our attention on the area- entropy law of {\it macroscopic} black holes  will be reconsidered in the final section from the viewpoint of   {\it the limits of applicability} of Bekenstein-Hawking's  Area-Entropy Law, on the basis of unified consideration of black holes ranging from the macroscopic to the extremely  {\it microscopic} scales given in the next section.

\section{\normalsize Entropy $S(V_3^L)$ and Mass $M ( V_3^L)$ of  $D_0$ brane gas systems ranging from Macroscopic to Microscopic scales}

\subsection { \normalsize {\bf KHR} in Yang's Quantized space-time with $d=3$}

Now, let us examine more in detail  |\bf KHR}  Eq.(2.1) together with Eq.(2.4),  specifically in $d=3.$  First of all, we notice  that there holds the following {\it simple} and {\it unified} expression
\begin{eqnarray}
{\rm n_ {dof}}\ (V_3^L)  =  dim\ ( \rho_{[L/\lambda]}\ (V_3^L)) =([L/\lambda]+1)^2
\end{eqnarray}
corresponding to Eq. (2.4) specifically in the case of  $ d=3,$ without any approximation and thus it enables us safely to investigate  the structure of the  {\it microscopic} system, together with the {\it macroscopic} system which was reviewed in the preceding section.

Here,  one should  notice very importantly that the above relation (3.1) manifestly shows``Kinematical reduction of spatial degrees of freedom"[10].  That is , ${\rm n_ {dof}}\ (V_3^L) (= ([L/\lambda]+1)^2 ) $ is not proportional to the order of  $V_3^L$ or $(L/\lambda)^3$ but proportional to $[(L/\lambda)]^2,$ remarkably  for the macroscopic system, on account of its own nature of underlying noncommutative quantized space and time, and leads us automatically to the Kinematical Holographic Relation {\bf KHR} as shown below  {\it without any recourse to the so-called Holographic Principle.}[21-22]

Indeed, in $d=3,$ the boundary area  ${\cal A}\ (V_3^L) $ is given by Eq. (2.5), that is,  ${\cal A}\ (V_3^L)  =  4\pi (L/\lambda)^2,$ so the relation (3.1) leads us to  the following form of the kinematical holographic relation  
\begin{eqnarray}
{\rm {\bf KHR}} \hspace{1cm} n_{\rm dof} (V_3^L) \cong  ( ({\cal A} (V_3^L) /4\pi)^{1/2} + 1)^2 , 
 \end{eqnarray}
taking into consideration the possible slight difference between $L/\lambda$ and $[L/\lambda]$. 

Meanwhile, as was remarked at the end of the preceding section and will be discussed in the subsection 5.2 in the final section, one should notice  that this form of 
{\bf KHR} Eq.(3.2) has a possibility of causing  {\it significant change}  of Area-Entropy Law of Black Holes ranging from macroscopic to microscopic scales, by rewriting Eq.(3.1) in the following form
\begin{eqnarray}
{\bf KHR}'  \hspace{1cm}
 {\rm n_ {dof}}\ (V_3^L)  \cong {\cal A} (V_3^L) /4\pi + 2 ({\cal A} (V_3^L) /4\pi)^{1/2} + 1,
\end{eqnarray} 
which surely reproduces the relation Eq. (2.1) with $f_3 \sim 4\pi $ given for the macroscopic system.    

\subsection{\normalsize Statistical derivation of   $S(V_3^L)$ and $M(V_3^L)$ based on {\bf KHR}}

According to I, let us consider the quantum system realized  in Yang's quantized space-time, which constitutes of $D_0$-branes   or D-particles [11-12].  As was done in I,  $D_0$ brane gas system formed inside $V_3^L$ is most likely described in terms of the {\it second-quantized field} of $D_0$ brane  or D-particle defined in Yang' quantized space-time,  $V_3^L$. 

Corresponding to Eq.(2.7), the representation space of  $  \rho_{[L/\lambda]}\ (V_3^L) $, called  ``Hilbert space I" in I (in distinction to ``Hilbert space II"), one finds 
\begin{eqnarray}
 \rho_{[L/\lambda]}\ (V_3^L): \quad |\ m \ \rangle,  m= 1,2,\cdots, ( [L/\lambda]+1)^2 (=n_{\rm dof}(V_3^L)) .     
\end{eqnarray} 
Needless to say, under this representation, each  spatial operators
 ${\hat X}_ i  's$ becomes expressed by
 $( [L/\lambda]+1)^2\times ( [L/\lambda]+1)^2$ matrix
like $\langle m\ |{\hat X_i } |\ n  \rangle$. 

Let us simply assume that the quantum states of $D_0$ branes of microscopic system are constructed in  ``Hilbert Space II," in a similar way,  as was done for the macroscopic system in I. Indeed, we assume that  even in the present {\it microscopic} system, $D_0$ brane gas model holds where all interactions among  $D_0$ branes are ignored so that the statistical operator at each $[site\ m]$,  ${\bf W}[m]$ is {\it common} to every [site] and given in the following form,
\begin{eqnarray} 
{\bf W}[m] = \sum_ k  w_k\  |\ [m]: k \rangle\  \langle k :[m]\ |,
\end{eqnarray}
with 
\begin{eqnarray}
 |\ [m]:  k \rangle  \equiv {1 \over \sqrt{k!}}({\bf a}_m^\dagger)^{k}|\ [m]:0\rangle.\end{eqnarray}
In the above expression,  $ {\bf a }_m^\dagger $ and $|\ [m]:  k \rangle \ ( k=0, 1, \cdots)$, respectively,  denote creation operator of D-particle at [site m]  ( see I, section 4.1) and  the normalized quantum-mechanical state in Hilbert space II with 
$k$ $D_0$ branes constructed by ${\bf a}_m^\dagger$ on $|\ [m]:0\rangle,$ i.e., the vacuum state of $[site\ m]$. Namely, we assume  that the   $D_0$ brane gas system is under a static and equilibrium state with temperature $T$ and the statistical operator at each $[site\ m]$ is common to every [site] with the common values $w_k$'s :
\begin{eqnarray}
 w_k = e^{-\mu k/T} / Z(T),
\end{eqnarray}
where
\begin{eqnarray}
Z(T) \equiv \sum_{k=0}^\infty e^{-\mu k/T} = 1 / (1- e^{-\mu / T})
\end{eqnarray}
and   $\mu$ denotes the average energy or effective mass of the individual $D_0$ brane in $V_3^L$. 

 The statistical operator of total system in $V_3^L$, ${\bf W}(V_3^L)$, is now given by  
\begin{eqnarray}
{\bf W}(V_3^L) = {\bf W}[1] \otimes {\bf W}[2] \otimes \cdots \otimes {\bf W}[([L/\lambda]+1)^2].
\end{eqnarray}

Consequently, one finds that the entropy and the energy or effective mass  of the total 
system,  $S(V_3^L)$ and $M(V_3^L)$ are respectively given by 
\begin{eqnarray}
 S(V_3^L) = - {\rm Tr}\ [{\bf W}(V_3^L\ {\rm ln} {\bf W}(V_3^L)]
 = n_{\rm dof} (V_3^L) S[site] 
\nonumber\\
= ([L/\lambda]+1)^2 S[site]
\end{eqnarray}
and
\begin{eqnarray}   
M (V_3^L) =n_{\rm dof} (V_3^L)  \mu {\bar N}[site] = ([L/\lambda]+1)^2 \mu {\bar N}[site]
\end{eqnarray}
corresponding to Eq.(32) and Eq.(33) in I, respectively. In the above expressions, $S[site]$ in Eq.(3.10) denotes the entropy of each [site] assumed to be common to  every [site] and is given by
\begin{eqnarray}
S[site] &&\equiv - \sum_k w_k\  {\rm \ln} w_k = {\mu {\bar N}[site] \over T}  + \ln  Z(T)
\nonumber \\
 &&= - \ln (1 - e^{- \mu /T}) + {\mu \over T}\ ( e^{\mu /T}- 1)^{-1},
\end{eqnarray}
and  ${\bar N}[site]$ in  Eq.(3.11) the average occupation number of $D_0$ brane at each $[site]$
\begin{eqnarray}
{\bar N}[site] \equiv \sum_k k w_k = (\ e^{\mu / T}- 1)^{-1}.
\end{eqnarray}

At the end of this subsection, let us notice the following two relations 
\begin{eqnarray}
 T =& \mu / \ln (1 + {\bar N}^{-1}[site]) 
\end{eqnarray}
and  
\begin{eqnarray}
S [site] = \ln (1+ \bar N[site]) + \bar N [site] \ln (1+ \bar N ^{-1}[site]),
\end{eqnarray}
which are simply derived from Eq.(3.12) and Eq.(3.13), respectively.

\section{Schwarzschild black holes ranging from Macroscopic to Extremely Microscopic scales}

Now, according to the consideration given in I, let us assume that the present $D_0$
 brane gas system ranging from Macroscopic- to Microscopic-scales in $ d=3,$ considered in the preceding subsection 3.2, transforms into a Schwarzschild black hole.  Indeed,  as was done in I,  we assume that the relevant quantities acquire certain limiting values, such like $ \mu_S$, ${\bar N}_S[site]$ and $S_S [site]$, while the size of the gas system, $L$, becomes $R_S$, th.at is, the so-called Schwarzschild radius given by
\begin{eqnarray}   
R_S =  2 G M_S(V_3^{R_S}) /c^2,
\end{eqnarray}
where $G$ and $c$ denote Newton's constant and the light velocity, respectively, and $M_S(V_3^{R_S})$ is given by Eq.(3.11) with $L= R_S$, $\mu = \mu_S$ and ${\bar N}[site] = {\bar N}_S [site].$  Indeed, inserting the above values into (3.11), we arrive at the important relation, called hereafter the black hole condition BHC, that is,
\begin{eqnarray}
{\rm BHC} \hspace{0.5cm} { M_S(V_3^{R_S}) \over  (2M_S(V_3^{R_S})/M_P +1)^2 }\
  (=  { M_S(V_3^{R_S}) \over n_{\rm dof}  (V_3^{R_S}) })
= \mu_S {\bar N}_S[site].
\end{eqnarray}
In the last expression, we assumed that $\lambda$, i.e., the short scale parameter in Yang' quantized space-time (see Appendix  A) is identified with Planck length $l_P = [G \hbar / c^3]^{1/2} = \hbar /( c M_P )$, where $M_P$  denotes Planck mass. In what follows, we will use Planck units in $D=4$ or $d=3$, with $M_P = l_P = \hbar = c = k =1,$ where $ k$ is Boltzmann's  constant [22].

According to the above consideration, let us notice $S_S(V_3^{R_S})$ given through Eq.(3.10), that is,
\begin{eqnarray}
 S_S(V_3^{R_S}) (= n_{\rm dof} (V_3^{R_S}) S_S[site]) = ([R_S/\lambda]+1)^2 S_S[site].
\end{eqnarray}

Furthermore, it is important here to notice that, by using Eq.(3.15), one finds the following relation
\begin{eqnarray} 
S_S[site]  = \ln (1+{ \bar N}_S [site]) +{ \bar N}_S  [site] \ln (1+ {\bar N}_S ^{-1}[site])
\end{eqnarray} 
which shows the fact that  $\bar N_S[site]$ gets some {\it universal} and fixed value, that is,
\begin{eqnarray}
 \bar N_S[site] \sim 1/ 0.12, 
\end{eqnarray}
under  $S_S [site] = \pi $  Eq.(2.10), that is, our basic assumption on  $S_S [site] = 4\pi\eta$ Eq.(2.9) with $\eta=1/4. $ 

\subsection {\normalsize Macroscopic black holes in d=3}
Now, let us notice that the BHC (4.2) becomes for the {\it macroscopic scales of black holes} $( M_S(V_3^{R_S}) \gg M_P)$
\begin{eqnarray}
{\rm BHC} \hspace{0.5cm}  M_S(V_3^{R_S}) = {M_P^2 \over 4 \mu_S {\bar N}_S[site]} 
\quad ({\rm for}\  M_S(V_3^{R_S}) \gg M_P), 
\end{eqnarray}
which reproduces Eq.(41) in I. 
 
On the other hand, the above relation (4.6) leads us to the following {\it universal} relation for the {\it macroscopic black holes} 
\begin{eqnarray} 
\mu_S M_S  \sim 0.03 \quad ({\rm for}\  M_S(V_3^{R_S}) \gg M_P)
\end{eqnarray}
in Planck units, on account of Eq.(4.5). 

\subsection {\normalsize Two kinds of Temperatures of Macroscopic black holes,  $T_{H.R.}$ and  $ T_S $ in d = 3}
As was once pointed out and argued in [12], one should notice here the fact that there exist two kinds of temperatures of black holes,  $T_{H.R.}$and $ T_S $.

The first one  $T_{H.R.}$ is the familiar Hawking's radiation temperature, which is given by using Eq. (4.3) in the following way:
\begin{eqnarray}
T_{H.R.}^{-1} = {d \over dM_S} S_S (V_3^{R_S}) =  {d \over dM_S}([R_S/\lambda]+1)^2 S_S[site] \\
\nonumber
=4\pi  (2M_S +1)
\end{eqnarray}
or
\begin{eqnarray}
T_{H.R.} = 1/(4\pi (2M_S +1)).
\end{eqnarray}

In the above derivation of Eq.(4.8) and (4.9), we assumed implicitly that $S_S[site]$ is independent of $M_S$  and  $S_S [site] =\pi$  according to the preceding arguments on the idea of {\it universality} of  $S_S [site]$, given in connection with Eq.(2.9) and Eq.(2.10). Further, one should notice that the relations Eq.(4.8)  and thus Eq.(4.9) are based on the relation $ S_S (V_3^{R_S}) =  ([R_S/\lambda]+1)^2 S_S[site]$ Eq.4.3). 
In this connection, we notice that   $T_{H.R.} = 1/(4\pi (2M_S +1))$ Eq.4.9) reproduces nicely the familiar result $T_{H.R.} = 1/ (8 \pi M_S)$ for the {\it macroscopic black holes } (see Eq.(48) in I).  On the other hand, however, it implies a possibility of causing the {\it nontrivial} modification for the {\it extremely microscopic} black holes, as will be shown in the next subsects. 4.3 and 4.4. 

 The second one $T_S$ is derived through Eq.3.14)
\begin{eqnarray}
 T_S =& {\mu}_S / \ln (1 + {\bar N_S}^{-1}[site])
\end{eqnarray}
or 
\begin{eqnarray}
 T_S =& {\mu}_S / \ln (1 +4 \mu_S M_S),
\end{eqnarray}
on account of Eq.(4.6).

For the macroscopic black holes, Eq.(4.9) and Eq.(4.7) show some {\it similarity} between the order of magnitudes of $T_{H.R.} \sim 1/ (8\pi M_S) \sim 0.04 /M_S$ and $\mu_S\sim 0.03/M_S$, that is, 
\begin{eqnarray}
T_{H.R.} \sim (0.04 / 0.03)\  \mu_S \sim 1.33 \mu_S.
\end{eqnarray} 
In contrast, one finds in this case 
\begin{eqnarray}
T_S \sim& (1 /0.11)\ \mu_S \sim 9.09\mu_S
\end{eqnarray}
from Eq.(4.10) or Eq.(4.11). The physical implication of Eq.(4.12) and Eq.4.13) will be discussed at the end of the subsection 5.1 in comparison with the corresponding  result of  the extremely microscopic black hole given in the next subsects 4.3 and 4.4. 

 \subsection{\normalsize Extremely  Microscopic black hole system in d=3} 
According  to the argument given in the beginning of this section, now we consider the extremely microscopic black hole system with $R_S =L = \lambda \ (= l_P) $ or $[L/\lambda] = 1$ in d=3. 

Let us denote hereafter the relevant quantities $M_S, T_{H.R.}, T_S$ and so on by attaching the tilde-mark such like ${\tilde M}_S,$ showing their specific values proper to the extremely microscopic system. 

First of all, in the extremely microscopic system, one finds  
\begin{eqnarray}
{\rm n_ {dof}}\ (V_3^\lambda)  =4,
\end{eqnarray}
from Eq.(3.1), and correspondingly
\begin{eqnarray}
{\tilde S}_S (V_3^\lambda) ( = n_{\rm dof} (V_3^\lambda) S_S [site]) = 4 S_S [site] ,
\end{eqnarray}
\begin{eqnarray}   
{\tilde M}_S (V_3^{R_S (= \lambda )}) (=n_{\rm dof} (V_3^\lambda) {\tilde \mu}_S {\bar N}_S [site])  =4 {\tilde {\mu}}_S {\bar N}_S [site]
\end{eqnarray}
to hold, from Eq.(3.10) and Eq.(3.11), respectively. The latter relation is further constrained from Eq. (4.1) as 

\begin{eqnarray}
{\tilde M}_S (V_3^{R_S (= \lambda )}) (= R_S/2 = \lambda /2) = 1/2
\end{eqnarray}
in Planck units.

Then, from Eq.(4.16) combined with Eq. (4.5) and Eq. (4.17), one finds 
\begin{eqnarray}
{\tilde \mu}_S (= {\tilde M}_S ( V_3^\lambda)/ (4{\bar N}_S[site]) ) = {1 \over 8} \times  0.12 \sim 0.02.
\end{eqnarray}

Finally, let us notice that  Eq.4.15)  tells us that the entropy of {\it extremely microscopic black hole } is given by
\begin{eqnarray}
 {\tilde S}_S(V_3^\lambda) ( = n_{\rm dof} (V_3^\lambda) S_S[site]) = 4 S_S[site] ) = 4\pi ,
\end{eqnarray}
under $S_S[site] ) = \pi$  Eq.(2.10).

\subsection{ Two kinds of Temperatures of Extremely Microscopic black hole,  ${\tilde T}_{H.R.}$ and ${\tilde T}_S$ in d=3 }

Corresponding to the arguments given in subsect. 4.2, let us consider two kinds of temperatures,  ${\tilde T}_{H.R.}$ and ${\tilde T}_S$ of extremely microscopic black hole system.

Firs of all, let us notice from Eq.(4.17)
\begin{eqnarray}
{\tilde M}_S =1/2.
\end{eqnarray}

 With respect to  ${\tilde T}_{H.R.}$,  as was emphasized in subsect. 4.2,  the expression $T_{H.R.} = 1/(4\pi (2M_S +1))$ given in Eq.(4.9) holds ranging from macroscopic to extremely microscopic system, so one immediately gets the following result
\begin{eqnarray}
{\tilde T}_{H.R.} (= 1/(4\pi (2{\tilde M}_S +1)) =1/(8\pi) \sim 0.04
\end{eqnarray}
on the basis of Eq(4.20).

On the other hand, according to Eq. (3.14), ${\tilde T}_S$ is simply given by  
\begin{eqnarray}
{\tilde T}_S = {\tilde \mu}_S / \ln (1 + {\bar N}_S^{-1}[site]).
\end{eqnarray}
By using ${\bar N}_S [site] \sim 1/0.12 $ Eq.(4.5), one gets
\begin{eqnarray}
{\tilde T}_S = {\tilde \mu}_S / \ln (1 + 0.12)  \sim (1/0.11) {\tilde \mu}_S \sim 9.09 {\tilde \mu}_S,
\end{eqnarray}
that is, the parallel result with Eq.(4.13). Further, by using the result Eq.(4.18),  one finally arrives at the result 
\begin{eqnarray}
{\tilde T}_S \sim 9.09 \times 0.02 \sim 0.18.
\end{eqnarray}

\section{\normalsize Concluding Arguments and Further Outlook}

\subsection{\normalsize Two kinds of Temperatures of black holes}

Let us reconsider the arguments about two kinds of temperatures of black holes given in subsects.4.2 and 4.4. First, concerning subsect. 4.2 devoted  to macroscopic black holes, we note the following three relations 
\begin{eqnarray}
 &T_S /T_{H.R.} ( \sim 9.09/1.33 ) \sim 6.83,  \nonumber \\
 &T_{H.R.}  ( \sim (0.04 / 0.03) ) \mu_S) \sim 1.33 \mu_S,\nonumber \\
 &T_S \sim 9.09\mu_S.
\end{eqnarray}
Next, concerning subsect. 4.4 devoted to the extremely microscopic black hole, we note the corresponding three relations
\begin{eqnarray}
 &{\tilde T}_S /  {\tilde T}_{H.R.} ( \sim 0.18/0.04) \sim 4.5,  \nonumber \\
 &{\tilde T}_{H.R.}  ( \sim  (0.04/0.02) {\tilde \mu}_S) \sim2.00 {\tilde \mu}_S,\nonumber \\
 &{\tilde T}_S \sim 9.09 {\tilde \mu}_S.
\end{eqnarray}

With respect to the marked difference between $T_S$ and $T_{H.R.}$ as seen in (5.1) and (5.2), one should notice that $T_S$ means the statistical and equilibrium temperature of $D_0$ brane gas (see Eq.(3.7)), that is, the temperature  {\it inside of black hole}, while  $T_{H.R.}$ is the thermodynamical temperature observed from {\it outside of black hole}.

We anticipate that the above arguments of possible existence and different behavior of two kinds of temperatures of black holes might be instructive for the forthcoming researches on formation and evaporation of black holes which may be closely related to the whole scales of black holes ranging from macroscopic to extremely microscopic.

\subsection{\normalsize Possible Modification of  Area-Entropy Law of Black Holes}

Finally, we reconsider our central concern,  the {\it universality }of Area-Entropy Law of Black Holes. By applying ${\bf KHR}'$ (3.3) to Eq.(3.10) which is derived through our simple D-particle gas model in section 3, we have
\begin{eqnarray}
 S(V_3^L)  \cong ( {\cal A} (V_3^L) /4\pi + 2 ({\cal A} (V_3^L) /4\pi)^{1/2} + 1 ) S[site].
\end{eqnarray}
For the {\it macroscopic} system, the first term on the right hand side becomes dominant term and the relation finally leads us to Bekenstein-Hawking area-entropy law of black holes under the assumption $S_S[site] =\pi $ Eq.(2.10).

On the other hand, one finds that for the {\it extremely microscopic} system, the relation  Eq.(5.3) just  leads us to Eq.(4.15) and Eq.(4.19) on account of ${\cal A} (V_3^\lambda) =4 \pi.$ This fact, however, implies very importantly that for the black holes of {\it intermediate} scales between macroscopic  and extremely microscopic, the second term on the right hand side of Eq.(5.3) has a possibility of causing the {\it significant correction term} proportional to $({\cal A} (V_3^{R_S}) /4\pi)^{1/2}$ to the familiar Bekenstein-Hawking's Area-Entropy Law. 

 We expect that such a possible {\it modification} of area-entropy law of black holes will shed a new light on the resolution of our serious question, Where does black hole entropy lie?[1] and related fundamental problems[6-9],[13-14].

As was remarked at the end in I [1], {\it Kinematical reduction of spatial degrees of freedom} which underlies {\bf KHR} may be expected to hold widely in the noncommutative space-time in general. Indeed, one can easily confirm that the original Snyder's quantized space-time  satisfies it  entirely in the same way as in the case of Yang's quantized space-time  shown in sect. 2 and sect. 3.

Before closing this short essay, let us note another interesting possibility of Yang's quantized space-time algebra (YSTA, see Appendix A). Indeed, one should notice that YSTA  is intrinsically equipped with the long scale parameter {\it R}, together with the short scale parameter $a$ which has been identified with Planck length in our present research so far. On the other hand, as was preliminarily pointed out in [25], ${\it R}$ might be promisingly related to a {\it fundamental cosmological constant} in connection with the recent dark-energy problem, under the further idea that YSTA subject to the {\it SO(D+1,1)} algebra (see, Appendix A) might be understood in terms of a some kind of local reference frame in the ultimate theory of quantum gravity, on the analogy of the familiar local Lorentz frame in  Einstein's General Theory of Relativity..

In this connection, we know that recently the issue of quantum space-time with nonvanishing cosmological constant has been addressed in the literature by several authors (See, for instance, [26-27]). It is quite interesting to examine their possible relations with our present approach.

We emphasize again the importance and the necessity of noncommutative geometry or more specifically Yang's quantized space-time towards ultimate theory of quantum gravity and Planck scale physics. It is our urgent task to {\it reconstruct} M-theory[20] in terms of noncommutative quantized space-time along this line of thought.

\vskip 1.5cm
\centerline {\Large  \bf Acknowledgments}
\vskip 0.8cm

The author would like to thank Shigefumi Mori, Masaki Kashiwara and Taichiro Kugo for valuable discussions and comments of {\bf KHR} over the arbitrary dimension $d.$ The author is grateful to Hideaki Aoyama for giving constant encouragement to the present research.

\vskip 1.5cm

\centerline{\Large\bf Appendix}
\vskip 1cm
\appendix
\section{ Yang's Lorentz covariant quantized space-time}
\label{appendixa}
Let us here briefly review the Lorentz-covariant Yang's quantized space-time [3, 4].  $D$-dimensional Yang's quantized space-time algebra (YSTA) was introduced  as the 
result of the so-called Inonu-Wigner's contraction procedure with two contraction parameters,  long $R$ and short $\lambda$\footnote {In the Yang's article [4], the short scale parameter is denoted by $a$ after the original Snyder's article [2].}, from $SO(D+1,1)$ algebra with generators $\hat{\Sigma}_{MN}$ [9] ( see, more in detail, [10] ) ; 

\begin{eqnarray} 
 \hat{\Sigma}_{MN}  \equiv i (q_M \partial /{\partial{q_N}}-q_N\partial/{\partial{q_M}}),
\label{sec_a: stp}
\end{eqnarray}
which work on $(D+2)$-dimensional parameter space  $q_M$ ($M= \mu,a,b)$ satisfying  
\begin{eqnarray}
             - q_0^2 + q_1^2 + \cdots + q_{D-1}^2 + q_a^2 + q_b^2 = R^2.
\end{eqnarray}
 
Here, $q_0 =-i q_D$ and $M = a, b$ denote two extra dimensions with space-like metric signature.

$D$-dimensional space-time and momentum operators, $\hat{X}_\mu$ and $\hat{P}_\mu$, 
with $\mu =1,2,\cdots,D,$ are defined in parallel by
\begin{eqnarray}
&&\hat{X}_\mu \equiv \lambda\ \hat{\Sigma}_{\mu a}
\\
&&\hat{P}_\mu \equiv \hbar /R \ \hat{\Sigma}_{\mu b},   
\end{eqnarray}
together with $D$-dimensional angular momentum operator $\hat{M}_{\mu \nu}$
\begin{eqnarray}
   \hat{M}_{\mu \nu} \equiv \hbar \hat{\Sigma}_{\mu \nu}
\end{eqnarray} 
and the so-called reciprocity operator
\begin{eqnarray}
    \hat{N}\equiv \lambda /R\ \hat{\Sigma}_{ab}.
\end{eqnarray}
Operators  $( \hat{X}_\mu, \hat{P}_\mu, \hat{M}_{\mu \nu}, \hat{N} )$ defined above 
satisfy the so-called contracted algebra of the original $SO(D+1,1)$, or YSTA :
\begin{eqnarray}
&&[ \hat{X}_\mu, \hat{X}_\nu ] = - i \lambda^2/\hbar \hat{M}_{\mu \nu}
\\
&&[\hat{P}_\mu,\hat{P}_\nu ] = - i\hbar / R^2\ \hat{M}_{\mu \nu}
\\
&&[\hat{X}_\mu, \hat{P}_\nu ] = - i \hbar \hat{N} \delta_{\mu \nu}
\\ 
&&[ \hat{N}, \hat{X}_\mu ] = - i \lambda^2 /\hbar  \hat{P}_\mu
\\
&&[ \hat{N}, \hat{P}_\mu ] =  i \hbar/ R^2\ \hat{X}_\mu,
\end{eqnarray}
with other familiar relations concerning  ${\hat M}_{\mu \nu}$'s omitted.

\section{Historical Background of Noncommutative Quantized Space and Time} 
\label{appendixb}

In association with the argument about Area-Entropy Law problem given in subsect. 5.2, let us consider another key problem towards ultimate theory of quantum gravity,  that is, ``Singularity Problem" in the local field theories, first disclosed  by Heisenberg-Pauli ($ \sim 1929)$.   In this connection, we recollect H.~Yukawa's ``Theory of Elementary Domain" (1966)   whose preliminary version, ``On Probability Amplitude in Relativistic Quantum Mechanics" (Talk in Japanese) started in the spring of 1934, stimulated by Dirac's idea of "Generalized Transformation Function" (g.t.f) (1933) presented in ``The Lagrangian in Quantum Mechanics,"[15]. It means that the above Yukawa's Talk was done  just in the midst of his  struggle with ``Meson Theory" (1934). Indeed, after one decade from ``Meson Theory,"the Dirac's idea ``g.t.f."was prominently referred in Yukawa's  elaborate work ``On the Foundation of the Theory of Fields" (1942) [16]. Furthermore, after the subsequent  ``Quantum  Theories of Non-local Fields" (1947 $\sim$ ), Yukawa  finally arrived at the thought of ``Atomistics and the Divisibility of Space and Time" (1966) [17] under a novel concept of  {\it Elementary Domain} $D,$ in association with the microscopic limit of Dirac's ``Generalized Transformation Function."

Yukawa's ``Theory of Elementary Domain" remained unaccomplished. However, he left  the following impressive statement (${\sim1978}$, in Japanese) [18]: ``When we will proceed in this direction, we shall be after all faced with the  problem of quantization of space-time $\cdots.$ The resolution, however, must be all left in future."

Nearly in the midst of 1990's, in accord with Yukawa's anticipation,  there appeared ``tantalizingly"[19]  {\it Noncommutative position coordinates of D-particles} in front of M-theory, that is, in quantum mechanics of many-body system of D-particles[19-20].  According to Yukawa's  viewpoint  on ``Second Quantization of Fields"[16], this fact strongly suggests the real existence of  {\it noncommutative quantized space-time}  behind D-particles or M-theory itself.  Motivated by this fact,  our present research started  in the form ``Space-time quantization and matrix model,"[23],\footnote{Let us note  that the central concept in our present approach,  ``Basis vector's set in Hilbert Space I,"that is,  $ |\ m\ \rangle 's \ (m=1, 2, \cdots, n_{\rm dof}) $ in Eq.(2.7) plays the role of Yukawa's ``Complete set of $D's."$ [17] } on the basis of the early works by H.S.~Snyder and C.N.~Yang (1942) [2-5], that is, ``Lorentz-covariant quantized space-time." The  historical background of their pioneering works in relation with W.K.~Heisenberg was referred in I, according to R.~Jackiw's comment [24]. 

Indeed, we  emphasize the historical importance of their pioneering  works, which will possibly play the ultimate role in clearing away {\it Twentieth-Century Clouds} over Difficulties of ultra-violet divergence in quantum field theories and Area-Entropy Law of black holes towards ultimate theory of quantum gravity and Planck scale physics [28].

\end{document}